\documentclass[journal]{IEEEtran}

\usepackage{amssymb}
\usepackage{amsmath}
\usepackage{amsfonts}
\usepackage{bbm}
\usepackage{braket}
\usepackage{algorithm}
\usepackage{algorithmicx}
\usepackage{theorem}
\usepackage{setspace}

\usepackage{url}
\usepackage{accents}
\usepackage{balance}
\usepackage{graphicx}
\usepackage{graphics}  
\usepackage{epsfig}    
\usepackage{color}
\usepackage{pgfplots}
\usepackage{epstopdf}
\usepackage{balance}

\newtheorem{theorem}{Theorem}
\newtheorem{remark}{Remark}
\newtheorem{lemma}{Lemma}

\newtheorem{proposition}{Proposition}

\usepackage{enumitem}
\setitemize[1]{itemsep=3.5pt,partopsep=0pt,parsep=\parskip,topsep=3.5pt}

\newcommand{\ubar}[1]{\underaccent{\bar}{#1}}

\DeclareMathOperator*{\argmin}{arg\,min}

\newcommand{\change}[1]{\textcolor{black}{#1}}
\newcommand{\StartChange}{\color{black}}
\newcommand{\EndChange}{\color{black}}

\begin{document}

\title{\bf \Large Distributed Optimization using ALADIN for
	\change{MPC} 
	   in Smart Grids
}
\author{Yuning~Jiang$^*$,
        Philipp~Sauerteig,
        Boris~Houska, 
        and~Karl~Worthmann
\thanks{$^*$Corresponding author.}
\thanks{Yuning Jiang and Boris Houska are with the School of Information Science and Technology, ShanghaiTech University, 201210 Shanghai, China  (e-mail: {\tt [jiangyn, borish]@shanghaitech.edu.cn}). Yuning Jiang is also with Chinese Academy of Sciences, Shanghai Institute of Microsystem and Information Technology, 200050 Shanghai, China and with the University of Chinese Academy of Sciences, 100049 Beijing, China.}
\thanks{Philipp Sauerteig and Karl Worthmann are with the institute for mathematics, Technische Universit\"at Ilmenau, 98693 Ilmenau, Germany (e-mail: {\tt [philipp.sauerteig, karl.worthmann]@tu-ilmenau.de}).} \thanks{Yuning Jiang and Boris Houska are supported by ShanghaiTech University, grant-no.\ F-0203-14-012. Philipp Sauerteig and Karl Worthmann are supported by the German Federal Ministry for Education and Research (BMBF; grant-no.\ 05M18SIA). Karl Worthmann is also indebted to the German Research Foundation (DFG-grant WO 2056/6-1).}}

\maketitle

\begin{abstract}
This paper presents a distributed optimization algorithm tailored \change{to solve} optimization problems arising in smart grids. 
In detail, we propose a variant of the Augmented Lagrangian based Alternating Direction Inexact Newton (ALADIN) method, 
which comes along with global convergence guarantees for the considered class of linear-quadratic optimization problems. 
We establish local quadratic convergence of the proposed scheme and elaborate its advantages compared to the Alternating Direction Method of Multipliers (ADMM). In particular, we \change{show} that, at the cost of more communication, ALADIN requires \change{fewer} iterations to achieve \change{the} desired accuracy. Furthermore, it is numerically demonstrated that the number of iterations is independent of the number of subsystems. 
The effectiveness of the proposed scheme is illustrated by running both an ALADIN and an ADMM based model predictive controller on a benchmark case study.
\end{abstract}


%
\IEEEpeerreviewmaketitle

\section{Introduction}
\label{sec:Intro}
The rapid uptake of renewable energy sources requires a fundamental transition of power networks from centralized to decentralized power generation~\cite{Ipakchi2009,Guo2013}. 
\change{Herein, locally distributed residential energy systems --~equipped with loads, generators, and energy storage devices~-- play a major role in order to successfully master the required paradigm shift. %
However, their incorporation in the electricity grid entails intermittent generation and bidirectional power flow and, thus, creates challenges for network power quality and stability~\cite{Olivares2014,Yu2016}. %
In particular, the volatile local energy generation results in peaks in the aggregated power demand profile, which can cause bottlenecks (even outages) or overload~\cite{Morstyn2018a}. %
Since the compensation of these fluctuations requires, in general, costly control energy, one of the grid operator's main objectives is flattening the aggregated power demand~\cite{Parhizi2015}, e.g., by using energy storage devices on a household level~\cite{Worthmann2014}. %
In~\cite{Telaretti2016a,Telaretti2016b,Shi2017} it is investigated how batteries can be exploited for peak shaving. %
Moreover, in~\cite{Brenna2016} batteries are used for dispatching photovoltaic power while in~\cite{Kohlhepp2019} thermal energy storages are considered. %
In practice, flexible coordination mechanisms are required in order to efficiently manage such systems despite the fact that the (local) battery dynamics are, in general, unknown to the grid operator}. %
Therefore, the associated optimization problems \change{have to} scale with the number of subsystems. 

Typically, problems \change{as described in the first paragraph} are embedded in a Model Predictive Control (MPC) framework~\cite{Parisio2014,Worthmann2014}, where at each time instant one optimization problem has to be solved. This makes the design of efficient online solvers inevitable. 
For this purpose, distributed optimization methods have been \change{developed to allow for parallelizable online calculations.} %
A classical approach is based on dual decomposition, where gradient-based first-order methods~\cite{Rantzer2009,Richter2011} are used to solve the concave dual problem. 
Alternatively, semi-smooth Newton methods~\cite{Frasch2015} can be applied.
However, such Newton-type methods are, in general, only convergent
if they are equipped with additional smoothing heuristics and
line-search routines. 
Compared to this, the Alternating Direction Method of Multipliers (ADMM)
has more favorable convergence \change{properties~\cite{Goldstein2014,Hong2017}.
Recently, many variants of ADMM were proposed that exploit inherent hierarchical structures.} %
One drawback of ADMM, however, is its scale dependency~\cite{Donoghue2016}, which is typically tackled via heuristic pre-conditioners in order to accelerate convergence. %
\change{Another way to construct a parallelizable online solver for MPC is to use classical Newton-type methods originally proposed for nonlinear programming such as Interior Point (IP) method~\cite{Necoara2009}, which can parallelize most of its operations. Although these methods converge much faster than ADMM and have the potential of parallel implementation, a global line-search routine is required to control the step size~\cite{Bitlisliouglu2017}.}
Recently, the Augmented Lagrangian based Alternating Direction
Inexact Newton (ALADIN) method \change{was} proposed in~\cite{Houska2016}. Similar to ADMM, it requires the local agents to solve small-scale decoupled problems and the central entity to solve a consensus problem in each iteration. 
\change{
However, in contrast to ADMM, ALADIN solves a coupled Quadratic Programming (QP) in the consensus step enabling locally quadratic convergence if suitable Hessian approximations are used.
}

\StartChange
For the particular problem of minimizing variability in power demand, a distributed algorithm was proposed in~\cite{Braun2016}. As a follow-up,~\cite{Braun2018} further exploited the available flexibility resulting from battery management and applied a centralized consensus variant of ADMM to increase efficiency and privacy. %
Note that, in this context distributed methods are required not just for computational efficiency, 
but even more importantly for maintaining privacy and enabling plug-and-play capability. In detail, the grid operator is not supposed to know the battery dynamics of the single households. %
However, due to its linear convergence~\cite{Hong2017}, the proposed ADMM variant requires many communication rounds between the grid operator and the residential energy systems to achieve a desired accuracy. To this end, Baumann et al. proposed to replace the expensive optimization routine by surrogate models in~\cite{Baumann2019}.

In this paper, we propose a tailored algorithm based on ALADIN for solving the optimization problem proposed in~\cite{Worthmann2015}, which aims at flattening the aggregated power demand profile in distribution grids. Due to the local quadratic convergence of the proposed algorithm 
under the assumption that the optimal solution is regular\footnote{\change{An optimal solution is called regular if the Linear Independence Constraint Qualification (LICQ), Strict Complementarity Conditions (SCC), and the Second Order Sufficient Conditions (SOSC) are satisfied~\cite[p.~591]{Nocedal2006}.}} as established in Theorem~\ref{thm::LocalConvergence}, the number of iterations and, thus, the total communication effort is reduced significantly. %
First, we reformulate the problem such that the objectives and inequality constraints are decoupled while only the affine equality constraints remain intertwined. 
Then, the proposed algorithm, similar to the standard ALADIN method~\cite{Houska2016}, alternates solving local small-scale problems and solving a coupled QP. However, our algorithm slacks the active local constraints in the coupled QP. Hence, the scale of the coupled QP is fixed. Moreover, we establish global convergence by using an $\ell_1$-penalty and provide guidelines on finding suitable tuning parameters. %
In addition, we numerically show that the required number of communication rounds is independent of the number of subsystems -- typically two are sufficient. Moreover, we design an MPC scheme based on the proposed algorithm and illustrate advantages compared to ADMM by applying it in a benchmark case study. 
\EndChange

The remainder is structured as follows: Section~\ref{sec:Model} recalls the physical model of residential energy systems and %
\change{a peak-shaving problem based on optimal energy storage control.} %
In Section~\ref{sec:ALADIN}, we propose the variant of ALADIN tailored to the structured optimization problem and \change{elaborate the implementation in detail. Moreover, we} establish local quadratic convergence as well as global convergence. 
\change{Then, ALADIN is embedded within an MPC scheme.} 
Based on a benchmark case study, the numerical results in Section~\ref{sec:comparison} show that ALADIN is scalable and outperforms ADMM. 

\change{
\textbf{Notations}
We use $(\mathbb{S}^n_+)$ $\mathbb{S}^n_{++}$ to denote the set of symmetric positive (semi-) definite matrices in $\mathbb{R}^{n\times n}$, $n \in \mathbb{N}$.} 
For a given $\Sigma\in\mathbb{S}_{++}^n$ and $x\in\mathbb{R}^n$, the notation
$
\|x\|_\Sigma^2\;=\;x^\top \Sigma x 
$
is used. 
\change{
For a number $p\geq 1$, the $p$-norm of $x\in\mathbb{R}^{n}$ is defined by
\[
\|x\|_p = \left(\sum_{i=1}^n|x_i|^p\right)^{\frac{1}{p}}\;\;\text{with}\;\;x=\begin{pmatrix}
x_1&\ldots&x_n
\end{pmatrix}^\top\;.
\]}%
Moreover, we use the notations
\[
\mathbbm{1}_n = \begin{pmatrix}
1& \ldots& 1
\end{pmatrix}^\top \in \mathbb{R}^n\;,\;0_n = \begin{pmatrix}
0& \ldots& 0
\end{pmatrix}^\top \in \mathbb{R}^n
\] 
and \change{denote the unit matrix in $\mathbb{R}^{n\times n}$ by $\mathbb{I}_n$}. For any integers $\ell \leq m$, we define $[\ell:m] := \{\ell,\ell+1, \ldots, m\} \subseteq \mathbb{Z}$.
\change{
The open ball with center $x\in\mathbb{R}^n$ and radius $r$ is denoted by $\mathcal{B}_r(x)=\{z\in\mathbb{R}^n\mid \|z-x\|_2<r \}$.
We call a function $f:\mathbb{R}^{n}\to\mathbb{R}\cup\{\infty \}$ strongly convex on a convex set $\Omega \subseteq \mathbb{R}^n$ with positive parameter $m>0$, if the inequality
\[
f(tx+(1-t)y) \leq tf(x) + (1-t)f(y) -\frac{m}{2}t(1-t)\left\|x-y\right\|_2^2
\]
holds for all $x,y \in \Omega$ and all $t \in [0,1]$.
The Kronecker product of two matrices $A\in\mathbb{R}^{m\times n}$ and $B\in\mathbb{R}^{p\times q}$ is defined by
\[
A\otimes B = 
\begin{bmatrix}
a_{11}B & \cdots & a_{1n}B\\
\vdots & \ddots & \vdots\\
a_{m1}B & \cdots & a_{mn}B
\end{bmatrix}
\in \mathbb{R}^{mp \times nq}\,.
\]
Finally, for a given function $f : \mathbb{R}^n \to \mathbb{R}$, we use the Landau notation 
\[
f(x) = \mathcal{O}(\left\| x \right\|)\;,\quad \text{if}\;\; \exists\, c \in \mathbb{R}\;,\;\lim_{x \to 0} \frac{f(x)}{\left\| x \right\|} = c\;.
\]}%

\section{System Model and Problem Formulation}
\label{sec:Model}
In this section, we recall the dynamic model of residential energy systems incorporating loads, energy generation, and storage devices~\cite{Braun2018}. 
Furthermore, we provide a mathematical formulation of possible goals of both the local systems and the grid operator, yielding an overall linear-quadratic optimization problem.

\subsection{Residential Energy Systems}
\label{sec:RES}
\begin{figure}[htbp!]
	\centering
	\includegraphics[scale=0.75]{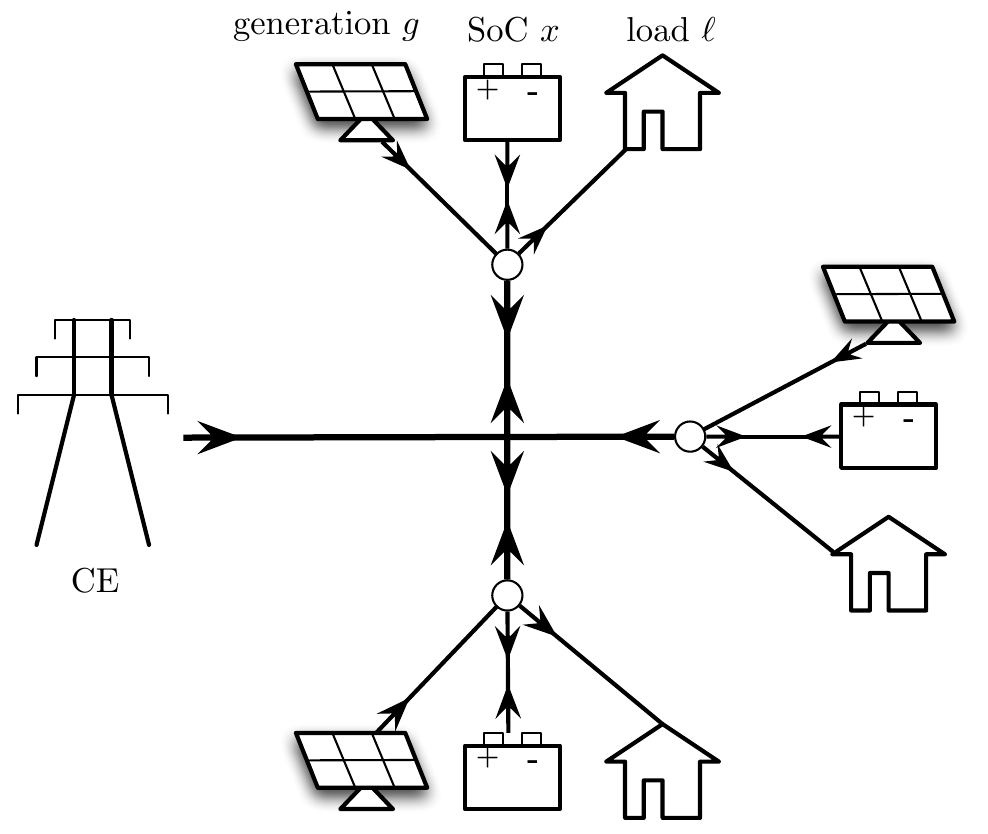}
	\caption{Electrical network consisting of several residential energy systems connected to a central entity.}
	\label{fig:network}
\end{figure}

In this paper, we consider a smart grid with $\mathcal{I} \in \mathbb{N}$ residential energy systems, which are coupled via a \change{Central Entity (CE)}, the grid operator\change{, see Figure~\ref{fig:network}}. The latter has to compensate the need for as well as the surplus of energy. %
\change{
Note that we are only interested in energy conservation on a residential level. We do not incorporate the grid topology. %
}

The battery dynamics of system~$i$, $i \in [1:\mathcal{I}]$, is given by
\begin{subequations}\label{subeq:dynRES}
	\begin{align}
	x_i(n+1) & =  \alpha_i x_i(n) + T(\beta_i u_i^+(n) + u_i^-(n)) \label{eq:dynRES1}\\[0.16cm]
	z_i(n) & =  w_i(n) + u_i^+(n) + \gamma_i u_i^-(n) \label{eq:dynRES2}
	\end{align}
\end{subequations}
with State of Charge (SoC) $x_i(n) \in \mathbb{R}$ of the battery in~kWh, power demand $z_i(n) \in \mathbb{R}$ in~kW at time instant $n \in \mathbb{N}_0$, and sampling time $T > 0$ in~h. The control inputs 
\[
u_i(n) = 
\begin{pmatrix} 
u_i^+(n)& u_i^-(n) 
\end{pmatrix}^\top 
\in \mathbb{R}^2
\]
represent the charging/discharging rate in~kW, while the net consumption $w_i(n) \in \mathbb{R}$ in~kW is load minus generation. The constants $\alpha_i, \beta_i, \gamma_i \in (0,1]$ model efficiencies with respect to self discharge and energy conversion, respectively. 

\change{At time~$k$, we consider the initial condition 
\[
x_i(k) = \hat{x}_i\;,\;i\in[1:\mathcal{I}],
\]
of the \mbox{$i$-th battery} with current measurements $\hat{x}_i$, and assume the future net consumption 
\[
w_i = \begin{pmatrix} 
w_i(k) & \cdots & w_i(k+N-1) 
\end{pmatrix}^\top
\] 
on the prediction window $[k:k+N-1]$}, $N \in \mathbb{N}_{\geq 2}$, to be given. 
The states and control inputs are subject to the constraints
\begin{subequations}\label{subeq:constraints}
	\begin{eqnarray}
	\label{eq:constraints:soc} 0 \quad \leq \quad & x_i(n) & \leq \quad C_i\,, \\
	\label{eq:constraints:discharge} \underline{u}_i \quad \leq \quad & u_i^-(n) & \leq \quad 0\,, \\
	\label{eq:constraints:charge} 0 \quad \leq \quad & u_i^+(n) & \leq \quad \overline{u}_i\,,\\
	\label{eq:constraints:chargeANDdischarge} 0 \quad \leq \quad & \frac{u_i^-(n)}{\underline{u}_i} + \frac{u_i^+(n)}{\overline{u}_i} \quad & \leq \quad 1\,.
	\end{eqnarray}
\end{subequations}
Here, $C_i \geq 0$ denotes the battery capacity of the $i$-th system, $\bar{u}_i$ and $\ubar{u}_i$ are the bound\change{s of the charging and discharging rates}, respectively. 
\StartChange
\begin{remark}
	Due to the recursive structure of the individual system \change{equation}~\eqref{eq:dynRES1}, the future \change{SoCs $x_i(k+n)$, $n\in[1:N]$,}
	are determined by~\eqref{eq:dynRES1} and the initial SoC $x_i(k)$, in particular \[
	\change{x_i(k\!+\!n) = \alpha_i^n \hat{x}_i + T \sum_{\ell=0}^{n-1} \alpha_i^{n-1-\ell} 
		\begin{pmatrix}
		\beta_i&1
		\end{pmatrix} u_i(\ell)}
	\]
	holds for all $i \in [1:\mathcal{I}]$ and $n \in [1:N]$. %
	Hence, state and input constraints~\eqref{subeq:constraints} of system~$i$ over the next~$N$ time steps can be summarized as polyhedral constraints 
	\begin{align}
	D_i u_i \; \leq \; d_i \notag
	\end{align}
	with matrix $D_i \in \mathbb{R}^{8N \times 2N}$ and vector $d_i \in \mathbb{R}^{8N}$, $i \in [1:\mathcal{I}]$.
\end{remark}
\EndChange
\subsection{\change{Objective: Load Shaping subject to Local Costs}}
\label{sec:Goal}
The grid operator is interested in flattening the aggregated power demand profile. One way is to track the overall net consumption $\change{\zeta} \in \mathbb{R}^N$, given by
\begin{equation}\notag
\change{\zeta(n) = \frac {1}{N} \sum_{j = n-N+1}^n \sum_{i=1}^{\mathcal{I}} w_i(j) }
\end{equation}
for $n \in [k:k+N-1]$, where \change{$k \geq N-1$} denotes the current time instant. The corresponding objective function can be modelled as $f_0: \mathbb{R}^N \to \mathbb{R}_{\geq 0}$,
\begin{equation}\label{eq::globalcost}
\begin{split}
f_0(\bar{z})&\;=\; \frac{\change{\sigma_0}}{N} \sum_{n=k}^{k+N-1} \left( \frac{\bar{z}(n)- \change{\zeta}(n)}{\mathcal{I}}  \right)^2 \\
&\;=\;  \frac{\sigma_0}{N\cdot \mathcal{I}^2}\left\|\bar{z}-\change{\zeta}\right\|_{2}^2
\end{split}
\end{equation}
with \change{$\sigma_0>0$} and $\bar{z}(n) =  \sum_{i=1}^{\mathcal{I}} z_i(n)$. 
In addition, we introduce the local costs $f_i:\mathbb{R}^{N}\times \mathbb{R}^{2N}\rightarrow \mathbb{R}_{\geq 0}$,
\begin{subequations}
	\label{eq::localcost}
	\begin{align}\label{eq::localcost1}
		f_i(u_i) & \;=\; \frac{\change{\sigma_i}}{2}\left(\left\| z_i - w_i\right\|_2^2 + \left\|u_i\right\|_2^2 \right)\\
		&\overset{\eqref{eq:dynRES2}}{=} \frac{\change{\sigma_i}}{2}\left(\left\|
		\left(\mathbb{I}_N \otimes \begin{pmatrix} 1 & \gamma_i \end{pmatrix}\right) u_i\right\|_2^2 + \left\|u_i\right\|_2^2\right)\\
		&\;=\;\frac{1}{2}\left\|u_i\right\|_{Q_i}^2
	\end{align}
\end{subequations}
with $\change{\sigma_i}\geq 0$,
$
z_i = \begin{pmatrix}
z_i(k)& \ldots & z_i(k+N-1)
\end{pmatrix}^\top,
$
and coefficient matrices
\begin{equation}\label{eq::Qi}
\begin{split}
Q_i
&\;=\;\change{\sigma_i} \left(\mathbb{I}_{2N} + \mathbb{I}_N \otimes \begin{bmatrix} 1& \gamma_i\\
\gamma_i&\gamma_i^2 \end{bmatrix} \right)
\end{split}
\end{equation}
for all $\;i \in [1:\mathcal{I}]$.
\change{As discussed in~\cite{Braun2018}, the first term on the right-hand side of~\eqref{eq::localcost1} ensures that each agent's output does not change drastically while the second term penalizes the charging and discharging effort. Moreover, the parameters~$\sigma_0$, $\sigma_i$, $i\in[1:\mathcal{I}]$, represent the weights of the global cost~\eqref{eq::globalcost} and local objectives~\eqref{eq::localcost}, respectively. %
The use of the \mbox{$i$-th} battery is penalized more 
with increasing~$\sigma_i$, $i \in [1:\mathcal{I}]$.} 

\subsection{Problem Formulation}\label{sec:ProbForm}
\change{The considerations in the previous subsections motivate} the optimization problem
\begin{subequations}
\label{eq::prob}
\begin{align}
	\min_{\bar{z},u}\quad &\;\;  f_0(\bar{z}) + \sum_{i=1}^{\mathcal{I}} f_i(u_i)\\[0.12cm]
	\text{s.t.}\quad &\;\; 
	\bar{z} = \overline{w}+\sum_{i=1}^{\mathcal{I}} A_iu_i  \qquad\mid \lambda\label{eq::coupled_equality}\\[0.12cm]
	%
	%
	&\;\;\change{ D_iu_i \leq d_i } \;,\quad i \in [1:\mathcal{I}], 
\end{align}
\end{subequations}
with $\overline{w} = \sum_{i=1}^{\mathcal{I}} w_i$. %
\change{The matrices $A_i \in \mathbb{R}^{N \times 2N}$ are given by
\begin{equation}\label{eq::Ai}
A_i = \mathbb{I}_N \otimes  \begin{pmatrix} 1 & \gamma_i  \end{pmatrix}\;,\;i\in[i:\mathcal{I}]\;.
\end{equation}
while constraint~\eqref{eq::coupled_equality} introduces the coupling between the CE and subsystems. 
\begin{remark}\label{rm::LICQ}
The $A_i$ expression~\eqref{eq::Ai}	shows that the affine equality constraint~\eqref{eq::coupled_equality} satisfies the LICQ condition, i.e., matrix 
\[
\begin{bmatrix}
-\mathbb{I}_N&A_1&\cdots&A_{\mathcal{I}}
\end{bmatrix}
\]
has full row rank.
\end{remark}
Throughout this paper, we write the Lagrangian multiplier right after the constraints such that the notation "$|\,\lambda$" denotes the Lagrangian multipliers of constraint~\eqref{eq::coupled_equality}.} 

\change{Next, we state sufficient conditions for uniqueness of the optimal solution of Problem~\eqref{eq::prob}.}
\begin{proposition}\label{thm::solution}
	Let \change{$\sigma_0 > 0$} and Problem~\eqref{eq::prob} be feasible. Then, the optimal power demand $\bar{z}^*$ and corresponding dual $\lambda^*$ are unique. If in addition \change{$\sigma_i>0$} for all $i \in [1:\mathcal{I}]$, then the optimal control $u^*$ of~\eqref{eq::prob} is also unique. 
\end{proposition}
\textbf{Proof.}
First, let $\change{\sigma_0}>0$. Due to the compactness of \change{the feasible set $\Set{u_i \in \mathbb{R}^{2N} | D_i u_i \leq d_i}$ of the $i$-th system}, the set 
\[
\overline{\mathbb{Z}} =\Set{ \bar{z}\in\mathbb{R}^N |
	\begin{matrix}
	\bar{z} = \overline{w}+\sum_{i=1}^{\mathcal{I}} A_iu_i \\[0.16cm]
	\change{D_iu_i \leq d_i} \;,\; i \in[i:\mathcal{I}]
	\end{matrix}}
\]
is compact, see e.g.~\cite{Rockafellar2015}. Since \change{$f_0$} is strongly convex, the optimal output~$\bar{z}^*$ corresponding to~\eqref{eq::prob} and, hence, the dual~$\lambda^*$ are unique. This proves the first assertion. %
\change{Now, let $\change{\sigma_i}>0$} for all $i \in [0:\mathcal{I}]$. Then, Problem~\eqref{eq::prob} is a strongly convex quadratic programming. Furthermore, the constraint Jacobian of~\eqref{eq::coupled_equality} has full row rank. Hence, the optimal solution $(\bar{z}^*,u^*,\lambda^*)$ is unique~\cite{Boyd2004}.  \hfill $\square$ 

\bigskip
\change{
From now on, we assume $\sigma_i > 0$ for all $i \in [1:\mathcal{I}]$, which implies that the matrices $Q_i$ are invertible.
Solving problem~\eqref{eq::prob} by using a centralized method assumes the existence of an all-knowing entity, i.e. the gird operator knows the dynamical model of each household as well as its predicted load and generation. This is impractical in case of several thousand households. Motivated by this, distributed or parallelizable methods~\cite{Worthmann2015,Braun2016,Braun2018} have been developed to solve~\eqref{eq::prob}. However, as discussed in Section~\ref{sec:Intro}, 
existing methods either have a slow convergence or require a global line-search routine, which needs many undesired communication rounds between the grid operator and the residential energy systems. In the next section, we propose a line-search free distributed method with fast local convergence to reduce the number of communication rounds significantly.
} 

\section{ALADIN for Smart Grids}
\label{sec:ALADIN}
In this section, we propose a variant of the ALADIN algorithm to solve~\eqref{eq::prob} in a distributed manner. Furthermore, we establish 
\change{both global convergence as well as
a locally quadratic convergence for the proposed algorithm} and show how it can be embedded efficiently in an MPC scheme.  

\subsection{\change{Distributed Optimization with ALADIN}}\label{subsec::aladin}
This section presents a variant of the ALADIN method to solve~\eqref{eq::prob}.  
Similar to the \change{standard} ALADIN method proposed in~\cite{Houska2016}, there are two main steps, a parallelizable step and a consensus step. 
\subsubsection{\change{Parallelizable Step}}
The small-scale QPs 
\begin{equation}\label{eq::deQP}
\begin{split}
\min_{v_i}\;&\;f_i(v_i)-(A_iv_i)^\top \lambda + \frac{1}{2}\left\|v_i-u_i\right\|_{Q_i}^2\\
\text{s.t.}\;\;&\;D_iv_i\leq d_i\qquad \mid\kappa_i
\end{split}
\end{equation}
are solved in parallel for all $i\in[1:\mathcal{I}]$. 
\change{Here, $(u_i, \lambda)$ denotes the current primal and dual iterates of ALADIN for solving~\ref{eq::prob}.}
If we denote the primal-dual solution of~\eqref{eq::deQP} by $(v_i,\kappa_i)$, the first-order optimality condition of~\eqref{eq::deQP} is given by
\[
0= \nabla f_i(v_i) + D_i^\top \kappa_i -A_i^\top \lambda + Q_i(v_i-u_i)\;.
\]
\change{
Hence, the modified gradient $g_i= \nabla_{u_i} \{f_i(v_i) + \kappa_i^\top D_i v_i\}$ can be evaluated by
\begin{equation}\label{eq::gradient}
g_i\;=\; A_i^\top \lambda + Q_i(u_i-v_i)
\end{equation}
in a derivative-free manner. As discussed in~\cite{Houska2016}, this parallelizable step exploits the distributed structure of~\eqref{eq::prob} by using dual decomposition and augmented Lagrangian}. 

\subsubsection{\change{Consensus Step}}
The following equality constrained QP is solved,
\StartChange
\begin{subequations}\label{eq::cQP}
	\begin{align}
	(\bar{z}^+,u^+,&s^+):=\argmin_{\bar{z},u,s}\quad  f_0(\bar{z})\\\notag
	&+ \sum_{i=1}^\mathcal{I}\left(\frac{1}{2}\|u_i - v_i\|_{Q_i}^2 + u_i^\top g_i + \frac{\mu_i}{2}\|s_i\|_2^2\right) \\
	\text{s.t.}\;\;\;&\,\bar{z} = \overline{w} + \sum_{i=1}^{\mathcal{I}} A_i u_i\qquad \mid\lambda^{+} \label{eq::coupledaffine}\\
	&\,D_i^\mathrm{act} (u_i -v_i) = s_i\;\;\quad \mid \kappa_i^\mathrm{QP}\;,\;i\in[1:\mathcal{I}]\;,\label{eq::decoupledaffine}
	\end{align} 	
\end{subequations}
where matrix $D_i^\mathrm{act}$ denotes the active Jacobian matrix at the local solution~$v_i$, i.e., $D_i^\mathrm{act}v_i = d_i^\mathrm{act}$ for all $i \in [1:\mathcal{I}]$. The solution $(u^+,\lambda^+)$ of~\eqref{eq::cQP} is used to update the primal and dual iterates. 
\EndChange
The value of the slackness parameters~$\mu_i$ reflects how much system~$i$ trusts the local active constraints. Later on, it is shown that updating~$\mu_i$ helps to improve the local convergence rate. 
\change{
\begin{remark}
Compared to the standard ALADIN~\cite{Houska2016}, the consensus QP~\eqref{eq::cQP} relaxes the decoupled equality constraints but not the coupled affine constraints. If we denote scaling matrices
\begin{equation}\label{eq::Hi}
H_i = Q_i + \mu_i {D_i^\mathrm{act}}^\top D_i^\mathrm{act}\;,\;i\in [1:\mathcal{I}],
\end{equation}
QP~\eqref{eq::cQP} can be rewritten in the form
\begin{subequations}\label{eq::cQP2}
	\begin{align}
	\min_{\bar{z},u}&\quad  f_0(\bar{z}) + \sum_{i=1}^{\mathcal{I}} \left(\frac{1}{2}\left\|u_i - v_i\right\|_{H_i}^2 + u_i^\top  g_i\right)\\
	\mathrm{s.t.}\;&\quad  \bar{z} = \overline{w} + \sum_{i=1}^{\mathcal{I}}A_i u_i  \qquad\mid \lambda\;.\label{eq::coupledaffine2}
	\end{align}
\end{subequations}
The advantage is that the scale of QP~\eqref{eq::cQP2} is fixed and does not depend on the number of active constraints. However, in contrast to the ADMM variant proposed in~\cite{Braun2018} the consensus QP~\eqref{eq::cQP2} is still large scale due to its dependency on the number of systems. \\
Moreover, as discussed in Remark~\ref{rm::LICQ},  constraint~\eqref{eq::coupledaffine2} satisfies the LICQ such that QP~\eqref{eq::cQP2} is always feasible independently of the feasibility of~\eqref{eq::prob}. %
Note that as the sum of a positive definite and a positive semi-definite matrix, $H_i$ is positive definite.
\end{remark}
}

\subsection{\change{Algorithm with Implementation Details}}\label{subsec::implementation}
\change{
In this section, we elaborate the implementation details, cp. Algorithm~\ref{alg:standard_ALADIN2}, of the distributed method proposed in Section~\ref{subsec::aladin}.}

\change{
Before we have a closer look at Algorithm~\ref{alg:standard_ALADIN2}, let us introduce two important ingredients used within the algorithm. First, since QP~\eqref{eq::cQP2} only incorporates equality constraints}, the analytical solution is given by 
\begin{subequations}
	\begin{align}
	\label{eq::dual_update}	
	\lambda^{+} \;=\;& \Lambda^{-1} \left(\zeta - \overline{w} +\sum_{i=1}^{\mathcal{I}} A_i(H_i^{-1}g_i - v_i) \right)\;,\\[0.12cm]
	\label{eq::primal_update}
	u_i^{+}\; =\;& v_i  + H_i^{-1}(A_i^\top \lambda^+ - g_i)\;,\;i\in[1:\mathcal{I}]\;.
	\end{align}
\end{subequations}
\change{Here, the scaling matrix $\Lambda$ is defined as
\begin{equation}\label{eq::dualHes}
\Lambda \;=\; \frac{N\cdot \mathcal{I}^2}{2\sigma_0}\mathbb{I}_N + \sum_{i=1}^{\mathcal{I}}A_iH_i^{-1}A_i^\top \;.
\end{equation}
The derivation is a direct consequence of the fact that the KKT conditions of~\eqref{eq::cQP2} are linear equations. 
Note that, the dual update~$\lambda^+$ requires communication between local subsystems and the CE while the primal update~$u_i^+$ is parallelizable.}   
As mentioned in Section~\ref{subsec::aladin}, in order to improve the local convergence, the parameters~$\mu_i$ need to be updated online. To this end, we introduce the \change{$\ell_1$-penalty based merit function $\Psi : \mathbb{R}^N \times \mathbb{R}^{2\mathcal{I} N} \to \mathbb{R}$,} 
\[
\Psi(\bar{z},u) \;=\; f_0(\bar{z})+\sum_{i=1}^{\mathcal{I}}f_i(u_i) + \bar{\lambda}\left\| \bar{z}-\overline{w}-\sum_{i=1}^{\mathcal{I}}A_iu_i\right\|_1
\]
with parameter $\bar{\lambda}>0$. %
\change{
Here, we assume that~$\bar{\lambda}$ is sufficiently large, which is important to establish the global convergence of Algorithm~\ref{alg:standard_ALADIN2} in Section~\ref{sec:Con}. 
Whenever, the merit function fulfils the descent condition~\eqref{eq::descent}, we set $\Pi = 1$, otherwise $\Pi = 0$, indicating the $\mu_i$ update. 
}%
\begin{algorithm}[hbp!]
	\StartChange
	\small
	\caption{A tailored ALADIN method for solving~\eqref{eq::prob}}
	\label{alg:standard_ALADIN2}
	\textbf{Initialization:}
	\begin{itemize}
		\item \textbf{Subsystems} $i\in[1:\mathcal{I}]$ do in parallel:
		\begin{itemize}
		\item[--] construct the local model as described in Section~\ref{sec:RES};
		\item[--] choose initial guess $u_i^0\in\mathbb{R}^{2N}$, and set $\mu_i=0$, $\ell=0$; 
		\item[--] send vectors $w_i$, $A_iu_i^0$ and scalar $\frac{1+\gamma_i^2}{\sigma_i(2+\gamma_i^2)}$ to the CE. 
		\end{itemize}
	
		\item \textbf{Central Entity} collects local information and does:
		\begin{itemize}
		\item[--] choose initial guess $\lambda^0\in\mathbb{R}^{N}$ and, send it to subsystems;
		
		\item[--] set terminal tolerance $\varepsilon>0$ and $\Pi = 0$;
		
		\item[--] compute $\zeta$, $\overline{w}$, and $\bar{z}^0=\overline{w}+\sum_{i=1}^{N}A_iu_i^0$;
		
		\item[--] precompute and save scaling matrix
		\begin{equation}\label{eq::Lambda}
		\Lambda_0^{-1} 
		\;=\; \left( 
		\frac{N\cdot \mathcal{I}^2}{2\sigma_0} + \sum_{i=1}^{\mathcal{I}}\frac{1+\gamma_i^2}{\sigma_i(2+\gamma_i^2)}
		\right)^{-1} \mathbb{I}_N\;.
		\end{equation}
		\end{itemize}
	
	\end{itemize}
	\textbf{Repeat:}
	\begin{itemize}
		\item[1)] \textbf{Subsystems} $i\in[1:\mathcal{I}]$ do in parallel: 
		\begin{itemize}
			\item[a)] If $\ell > 0$, 
			\begin{itemize}
				\item[--] receive $\lambda^{\ell+1}$ and $\Pi$ from CE;
				\item[--] compute 
				\[
				u_i^{\ell+1} =
				\left\{
				\begin{split}
				 v_i^\ell  + H_i^{-1}(A_i^\top \lambda^{\ell+1} - g_i^\ell)&\quad \text{if }\Pi = 1\;,
				\\[0.12cm]
				v_i^\ell  + Q_i^{-1}(A_i^\top \lambda^{\ell+1} - g_i^\ell)&\quad \text{if }\Pi = 0\;;
				\end{split}
				\right.
				\]
				
				\item[--] set $\ell\leftarrow \ell+1$.
			\end{itemize} 
			\item[b)] Compute
			\[
			\begin{split}
			v_i^{\ell}=\argmin_{v_i}&\;\;f_i(v_i)-  (A_iv_i)^\top \lambda^\ell + \frac{1}{2}\left\|v_i-u_i^{\ell}\right\|_{Q_i}^2\\
			\text{s.t.}&\;\;D_iv_i\leq d_i\qquad \mid\kappa_i
			\end{split}
			\]
			and set 
			$
			g_i^{\ell} = A_i^\top \lambda^\ell + Q_i(u_i^{\ell}-v_i^{\ell})
			$.	
			\item[c)] Set 
			\begin{equation}\label{eq::mu_update}
			\mu_i\;=\;\frac{\left\| \kappa_i \right\|_1}{\left\| D_i (v^{\ell}_i-u^{\ell}_i) \right\|_1}
			\end{equation}				
			and $H_i= Q_i + \mu_i {D_i^{\mathrm{act}}}^\top D_i^{\mathrm{act}}$.
			\item[d)] Send symmetric matrix $A_iH_i^{-1}A_i^\top$, vectors
			\[
			A_iv_i^{\ell}\;,\;c_{i,1} = A_i(H_i^{-1}g_i^{\ell} - v_i^{\ell})\;,\;c_{i,2}=A_i(Q_i^{-1}g_i^{\ell} - v_i^{\ell})
			\] 
			and scalars $f_i(v_i^\ell)$, $\delta_i=\|v_i^{\ell}-u_i^\ell\|_1$ to the CE.
		\end{itemize}
		
		\item[2)] \textbf{Central Entity} collects local information and does:
		\begin{itemize}
			\item[a)] If $\max_{i}\delta_i<\varepsilon$, terminate the algorithm;
			\item[b)] If $\ell = 0$, set $\psi = \Psi(\bar{z}^\ell,v^{\ell})$. Otherwise,  
			\begin{itemize}
				\item[--] if 
				\begin{equation}\label{eq::descent}
					\Psi(\bar{z}^\ell,v^{\ell}) < \psi\;,
				\end{equation}
				 update $\psi = \Psi(\bar{z}^\ell,v^{\ell})$, set $\Pi = 1$ and compute $\Lambda^{-1}$ by~\eqref{eq::dualHes};
				\item[--] else, set $\Pi=0$.
			\end{itemize}
			\item[c)] Compute the dual update $\lambda^{\ell+1}$ by 
			\[
			\lambda^{\ell+1} =\left\{
			\begin{split}
			 \Lambda^{-1} \left(\zeta - \overline{w} +\sum_{i=1}^{\mathcal{I}} c_{i,1} \right)&\quad \text{if }\Pi = 1\;,\\[0.12cm]
			\Lambda_0^{-1} \left(\zeta - \overline{w} +\sum_{i=1}^{\mathcal{I}} c_{i,2} \right)&\quad \text{if }\Pi = 0\;. 
			\end{split}
			\right.			
			\]
			\item[d)] Send $(\lambda^{\ell+1},\Pi)$ to subsystems and set $\bar{z}^{\ell+1} = \zeta- \frac{N\cdot\mathcal{I}^2}{2\sigma_0}\lambda^{\ell + 1}$.
		\end{itemize}
	\end{itemize}
	\EndChange
\end{algorithm}

\StartChange
In Algorithm~\ref{alg:standard_ALADIN2}, the integer~$\ell$ denotes the index of the iteration. During the initialization phase, the CE precomputes the scaling matrix $\Lambda_0^{-1}$ similar to~\eqref{eq::dualHes} with $\mu_i=0$, i.e. $H_i = Q_i$. Here, we use the Sherman-Morrison-Woodbury formula~\cite[Appendix~A]{Nocedal2006} to compute~$Q_i^{-1}$, 
\begin{equation}
\label{eq::invQi}
\begin{split}
Q_i^{-1}&\; =\; \frac{1}{\sigma_i} 
\left(
\mathbb{I}_{2N} - \frac{1}{2+\gamma_i^2}\mathbb{I}_N \otimes \begin{bmatrix} 1& \gamma_i\\
\gamma_i&\gamma_i^2 \end{bmatrix}
\right)\\[0.16cm]
&\;=\;\frac{1}{\sigma_i} \mathbb{I}_N\otimes
\left(
\mathbb{I}_{2} - \frac{1}{2+\gamma_i^2}\begin{bmatrix} 1& \gamma_i\\
\gamma_i&\gamma_i^2 \end{bmatrix}
\right)\\[0.16cm]
&\;=\;\frac{1}{\sigma_i(2+\gamma_i^2)} \mathbb{I}_N\otimes
\begin{bmatrix}
1+\gamma_i^2&-\gamma_i\\
-\gamma_i& 2
\end{bmatrix},
\end{split}
\end{equation}
which yields
\begin{equation}
\label{eq::dualBlock}
\begin{split}
A_iQ_i^{-1}A_i^\top &\;=\; \mathbb{I}_N\otimes \frac{
	\begin{pmatrix}
	1\\\gamma_i
	\end{pmatrix}^\top
	\begin{bmatrix}
	1+\gamma_i^2&-\gamma_i\\
	-\gamma_i& 2
	\end{bmatrix}
	\begin{pmatrix}
	1\\\gamma_i
	\end{pmatrix}
}{\sigma_i(2+\gamma_i^2)}\\[0.16cm]
&\;=\;\frac{1+\gamma_i^2}{\sigma_i(2+\gamma_i^2)}
\mathbb{I}_N\;. 
\end{split}
\end{equation}
This indicates that computing $\Lambda_0^{-1}$ by~\eqref{eq::Lambda} only requires the subsystems to send scalars 
$$\frac{1+\gamma_i^2}{\sigma_i(2+\gamma_i^2)}\;,\;i\in[1:\mathcal{I}]$$ 
to the CE. 

In the main loop, the subsystems first update $u_i^{\ell+1}$ in parallel based on the current~$\mu_i$ in Step~1a). 
Once the local QP~\eqref{eq::deQP} is solved and the gradient~$g_i^{\ell}$ is evaluated in Step~1b), parameter~$\mu_i$ and matrix~$H_i$ are computed in Step~1c). Then, the subsystems send the local information constructed in Step~1d) to the CE. Here, the symmetric matrix $A_iH_i^{-1}A_i^\top$ can be computed as
\begin{equation}\label{eq::dualHi}
\begin{split}
A_i&H_i^{-1}A_i^\top = \frac{1+\gamma_i^2}{\sigma_i(2+\gamma_i^2)}\cdot
\mathbb{I}_N - \frac{\mu_i}{\sigma_i^2(2+\gamma_i^2)^2}\\
&\cdot A_i{D_i^\mathrm{act}}^\top(\mathbb{I}_{n_i^\mathrm{act}} + \mu {D_i^\mathrm{act}}Q_i^{-1} {D_i^\mathrm{act}}^\top)^{-1}D_i^\mathrm{act}A_i^\top,
\end{split}
\end{equation}
where $n_i^\mathrm{act}$ denotes the number of rows of $D_i^\mathrm{act}$.
The explicit form~\eqref{eq::dualHi} implies that sending $A_iH_i^{-1}A_i^\top$ to the CE only requires $N\cdot n_i^\mathrm{act}$ floats.

After the CE collects the local information, Step~2a) checks the terminal condition of Algorithm~\ref{alg:standard_ALADIN2}. If 
\[
\max_{i}\;\;\|v_i^\ell - u_i^\ell\|_1 < \varepsilon
\]
holds, the current iterate~$v^\ell$ satisfies the stationary condition and primal condition of~\eqref{eq::prob} up to an error of order $\mathcal{O}(\varepsilon)$, i.e.,
\[
\begin{split}
&\sum_{i=1}^{\mathcal{I}}A_iv_i^\ell -\bar{z}^\ell - \overline{w}=\mathcal{O}(\varepsilon)\;,\\
\text{and}\quad &
\; g_i^\ell +A_i^\top\lambda^\ell = \mathcal{O}(\varepsilon),\;i\in[1:\mathcal{I}]\;.
\end{split}
\]
Otherwise, Step~2b) is executed based on the merit function~$\Psi$. 
In order to check the strict descent condition~\eqref{eq::descent}, the Armijo conditions~\cite[p.~540]{Nocedal2006} defined by
\begin{equation}\label{eq::Armijo}
\Psi(\bar{z}^{\ell},v^{\ell})\leq \psi + \eta\mathcal{D} \Psi (\Delta \bar{z}, \Delta v)\;\;\text{ with }\;\;\eta \in (0,1).
\end{equation}
is used. Here, $(\Delta \bar{z}, \Delta v)$ denotes the difference between the current iterates $(\bar{z}^{\ell},v^{\ell})$ and the previous iterates associated to $\psi$, and $\mathcal{D} \Psi(\Delta \bar{z}, \Delta v)$ denotes the directional derivative of~$\Psi$ in the direction $(\Delta \bar{z}, \Delta v)$. How to compute $\mathcal{D} \Psi(\Delta \bar{z}, \Delta v)$ has been elaborated in~\cite[Thm.~18.2]{Nocedal2006}. 
Then, in Step~2c) the dual iterate is updated and $(\lambda^{\ell+1},\Pi)$ is sent to the local systems in Step~2d).
\begin{remark}
	In practice, Armjio condition~\eqref{eq::Armijo} is expensive to check as it requires to compute $\mathcal{D} \Psi(\Delta \bar{z}, \Delta v)$. 
	One practical way to implement the descent condition~\eqref{eq::descent} is to check 
	\[
	\Psi(\bar{z}^\ell,v^{\ell}) \leq  \psi - \hat{\varepsilon}
	\]
	for a small constant $\hat{\varepsilon}$ of order of the machine precision (or local QP solver's accuracy). Although this can be interpreted as the weakest possible descent condition that can be checked with finite precision arithmetic, it works well in practice.
\end{remark}
\begin{remark}
	The merit function $\Psi$ only aims to control the update frequency of $\mu_i$. It never rejects steps or slows down the progress of the iterations in Algorithm~\ref{alg:standard_ALADIN2}, which is different from globalization routines such as line search methods when using centralized approaches.
\end{remark}
\EndChange

\subsection{Convergence Analysis}\label{sec:Con}
In this section, we analyze the \change{theoretical} convergence \change{properties} of Algorithm~\ref{alg:standard_ALADIN2} in two steps: local and global convergence. Since $\sigma_i>0$ for all $i\in[1:\mathcal{I}]$, \change{Proposition}~\ref{thm::solution} guarantees uniqueness of the optimal primal $(\bar{z}^*,u^*)$ and dual solution~$\lambda^*$ of~\eqref{eq::prob}.
\change{Moreover, we assume that Algorithm~\ref{alg:standard_ALADIN2} checks the Armijo condition~\eqref{eq::Armijo} at Step 2b) to determine the update of $\mu_i$.}
\change{
\begin{proposition}\label{thm::LocalConvergence}
	Let the optimal solution of Problem~\eqref{eq::prob} be regular and let $\mu_i$ be updated such that 
	\begin{equation}\label{eq::local_cond_mu}
	\frac{1}{\mu_i} < \mathcal{O} (\left\| v_i^\ell-u_i^\ell \right\|_2)\;,\;i\in[1:\mathcal{I}]  \;.
	\end{equation}
	Then, there exist constants $r,\eta > 0$ with $r \eta \leq 1$ such that all $v^\ell \in \mathcal{B}_{r}(u^*)$ satisfy
	\[
	\|v^{\ell+1}-u^*\|_2\leq \eta \|v^{\ell}-u^*\|_2^2 \;.
	\]
\end{proposition}
\textbf{Proof.  }
Locally, following the regularity assumption the active sets at the local solutions $v_i^\ell\in\mathcal{B}_r(u^*)$ are fixed such that the local solution $v^\ell$ of~\eqref{eq::deQP} can be represented as the image of $(u,\lambda)$ under an affine map, i.e., there exist some matrices $T_1,T_2 \in \mathbb{R}^{\mathcal{I} N \times \mathcal{I} N}$ such that
\begin{equation}
\label{eq::local_deQP}
v^{\ell} - u^* = T_1 (u^\ell-u^*)+ T_2(\lambda^\ell-\lambda^*) \;.
\end{equation}
Moreover, if the $\mu_i$'s satisfy the conditions~\eqref{eq::local_cond_mu}, the coupled QP~\eqref{eq::cQP2} yields the quadratic contractions
\begin{equation}
\label{eq::local_quadratic}
\begin{split}
\|u^{\ell+1} - u^*\|_2&\leq \frac{\alpha}{2}\|v^{\ell}-u^*\|_2^2\,,\\ \text{and}\quad \|\lambda^{\ell+1}-\lambda^*\|_2 &\leq  \frac{\alpha}{2}\|v^{\ell}-u^*\|_2^2
\end{split}
\end{equation}
with a positive constant~$\alpha$ similar to the Newton-type methods~\cite[Sec.~7]{Houska2016}.} Considering two consecutive iterations in Algorithm~\ref{alg:standard_ALADIN2}, we get
\[
\begin{split}
\|v^{\ell+1} - u^*\|_2 
\overset{\eqref{eq::local_deQP}}{\leq}\;\;& \|T_1\|_2 \|u^{\ell+1}-u^*\|_2+ \|T_2\|_2\|\lambda^{\ell+1}-\lambda^*\|_2 \\
\overset{\eqref{eq::local_quadratic}}{\leq}\;\;&\frac{\alpha}{2}(\|T_1\|_2+\|T_2\|_2) \|v^{\ell}-u^*\|_2^2\;,
\end{split}
\]
which concludes the proof with $\eta = \frac{\alpha}{2}(\|T_1\|_2+\|T_2\|_2)$. \hfill$\square$

\bigskip
\change{
This proposition establishes the local quadratic convergence rate of the iterates $v^\ell$ while implying that the local convergence progress benefits from the $\mu$ update.} %
\begin{remark}
	\change{
	If the iterates~$(v^\ell)_{\ell \in \mathbb{N}}$ converge locally with quadratic rate as discussed in Proposition~\ref{thm::LocalConvergence}, the descent condition~\eqref{eq::Armijo} always holds in a neighborhood of a regular minimizer~\cite{Nocedal2006}. Hence, $\mu_i$ is updated in every step. In Algorithm~\ref{alg:standard_ALADIN2}, the $\mu_i$ update is based on~\eqref{eq::mu_update} ensuring that~\eqref{eq::local_cond_mu} holds locally.}
\end{remark}

Concerning the global convergence analysis, we introduce an auxiliary function~\change{$\mathcal{L} : \mathbb{R}^{2\mathcal{I} N} \times \mathbb{R}^{N} \to \mathbb{R}$},
\[
\mathcal{L}(u,\lambda) =\left\Vert\lambda-\lambda^*\right\Vert_{\Lambda_0}^2+ \sum_{i=1}^{\mathcal{I}}\left\|u_i-u_i^*\right\|_{Q_i}^2 \;,
\]
to measure the distance of iterates~$(u,\lambda)$ to the optimum~$(u^*,\lambda^*)$. In order to establish global convergence we introduce the following technical result.
\begin{lemma}\label{lem::GlobalConvergence}
	Let Problem~\eqref{eq::prob} be feasible and $\mu = 0$ \change{at the current iteration of} Algorithm~\ref{alg:standard_ALADIN2}. Then, there exists a real number $m > 0$ such that the iterates satisfy the inequality
	\begin{equation}\label{eq::descentCond}
	\mathcal{L}(u^{\ell+1},\lambda^{\ell+1}) \leq \mathcal{L}(u^{\ell},\lambda^{\ell}) -4m \|v^{\ell+1} - u^*\|_2^2\;.
	\end{equation}
\end{lemma}
\change{
\textbf{Proof.  }
According to Proposition~\ref{thm::solution}, the choice of $\sigma_i>0$ for all $i\in[1:\mathcal{I}]$ indicates Problem~\eqref{eq::prob} is strongly convex. Then, the proof can be established following the proof of Theorem~1 in~\cite{Jiang2019} step by step. \hfill$\square$}

\bigskip
\change{
Since function $\mathcal{L}$ is bounded from below by $0$,  
Lemma~\ref{lem::GlobalConvergence} indicates that the sequence $(\mathcal{L}(u^{\ell},\lambda^{\ell}))_{\ell \in \mathbb{N}}$ converges as the algorithm progresses, i.e.~$v^{\ell}$ converges to the optimal solution if $\mu_i=0$ is fixed. However, this is in conflict with the local quadratic convergence established in Proposition~\ref{thm::LocalConvergence}, which requires to update~$\mu_i$. In order to overcome this problem and establish the global convergence of Algorithm~\ref{alg:standard_ALADIN2} with optionally updating~$\mu_i$, we introduce the following technical result.
\begin{proposition}
	\label{prop::exactness}
	Let the optimal solution~$(\bar{z}^*,u^*,\lambda^*)$ of Problem~\eqref{eq::prob} be regular and consider
\begin{align}\label{eq::merit_problem}
	\min_{\bar{z},u} \quad \Psi(\bar{z},u)\qquad \mathrm{s.t.}\quad 
	D_iu_i\leq d_i\;,\;i\in[1:\mathcal{I}]\;.
\end{align} 
If~$\bar{\lambda}$ satisfies
	\[
	\bar{\lambda} > \|\lambda^*\|_\infty = \max_{i}\;|\lambda_i^*|
	\]
	with $\lambda_i^*$ the $i$-th element of $\lambda^*$, then 
\begin{align}
	(\bar{z}^*,u^*) \text{ primal solution of~\eqref{eq::prob}} \Leftrightarrow (\bar{z}^*,u^*) \text{ solves~\eqref{eq::merit_problem}} \notag
\end{align}
holds, i.e. $\Psi$ is an exact merit function of Problem~\eqref{eq::prob}.
\end{proposition}
For details of the proof we refer to~\cite[Thm.~17.3]{Nocedal2006}. As $\bar{\lambda}$ is assumed to be sufficiently large, Proposition~\ref{prop::exactness} indicates the exactness of $\Psi$. Therefore, if the Armijo condition~\eqref{eq::Armijo} holds, the convergence of $v^{\ell}$ can be preserved with $\mu_i$~update. This result is summarized in the following theorem.}
\begin{theorem}\label{thm::GlobalConvergence}
	Let \change{$\sigma_i>0$, $i\in[0:\mathcal{I}]$} and Problem~\eqref{eq::prob} be feasible. \change{If the Armijo condition~\eqref{eq::Armijo} is used in Step~2d) of Algorithm~\ref{alg:standard_ALADIN2}}, the iterates $(v^\ell)_{\ell \in \mathbb{N}}$ in Algorithm~\ref{alg:standard_ALADIN2} converge globally to the optimal solution~$u^*$. If additionally the optimal solution of~\eqref{eq::prob} is regular, the iterates $v^\ell$ locally converge with quadratic rate.
\end{theorem}
\change{ 
\textbf{Proof.  }
The first statement follows Lemma~\ref{lem::GlobalConvergence} and the fact that~$\Psi$ is an exact merit function. As Proposition~\ref{thm::LocalConvergence} presented, the second statement holds since $v^\ell$ converges with quadratic rate after entering the local neighborhood~$\mathcal{B}_r(u^*)$. 
\hfill$\square$}
\begin{remark}
	In practice, the rigorous construction of the parameter $\bar{\lambda}<\infty$ in function~$\Psi$ requires meta-data from the user in order to ensure that~$\Psi$ is an exact merit function. However, state-of-the-art Sequential Quadratic Programming (SQP) solvers~\cite{Nocedal2006} implement effective heuristics for choosing~$\bar{\lambda}$ online in order to avoid this need for meta-data. In this paper, we update $\bar{\lambda}$ based on the heuristic used in the open source toolkit \texttt{ACADO}~\cite{Houska2011}. 
	\change{
		If the descent condition of $\Psi$ at Step~4b) holds, we update}%
	\[
	\change{\bar{\lambda} = 10\|\lambda^{\ell}\|_\infty\;.}
	\]
\end{remark}

\subsection{Distributed MPC Scheme Using ALADIN}
\label{sec:MPC}
In the context of MPC, Problem~\eqref{eq::prob} is solved during each sampling time step~$k$ based on the current measurements and then, each subsystem implements the first element of its control sequence, $u_i^*(k)$, $i\in[1:\mathcal{I}]$. This process defines an implicit feedback law that is applied iteratively~\cite{Worthmann2015}. 

Algorithm~\ref{alg::MPC} outlines an ALADIN-based distributed MPC scheme for smart grids.
Similar to the proposed ADMM-based MPC scheme in~\cite{Braun2018}, Algorithm~\ref{alg::MPC} does not assume the existence of some all-knowing entity. Each local agent only knows its own parameters and optimization problem. %
\change{
The CE collects only specific local information as outlined in Algorithm~\ref{alg:standard_ALADIN2} 
and broadcasts the updated dual variable as well as one float encoding whether~$\mu_i$ is updated or not. 
Furthermore, the control is initialized based on the optimal solution of the previous MPC. 
}

Although the reference trajectory in our setting is based on the (predicted) future net consumption, one could consider~$\change{\zeta} = \{\zeta(n)\}_{n\in[k:k+N-1]}$ at time step~$k$ as a segment of a \change{given sequence} such that we do not have to compute it in each MPC step. In that case, the reference does not depend on predicted values. \change{As a} result, the subsystems do not need to send~$w_i$ to the CE. 
\change{
Notice that the online choice of the reference does not affect the established convergence results applying for solving~\eqref{eq::prob}.
}

\begin{algorithm}[htbp!]
	\caption{ALADIN based distributed MPC scheme}
	\label{alg::MPC}
	{\bf Offline:}
	\begin{itemize}
		\item	Initial guess $(u,\lambda)$ and set $k=0$.
	\end{itemize}
	{\bf Online:}
	\begin{enumerate}
		\item \textbf{Subsystems} measure current SoC $x_i(k)$, predict future net consumption $w_i$ and send it to CE.
		\item  \textbf{Central Entity} computes the reference trajectory $\zeta$.
		\item Run Algorithm~\ref{alg:standard_ALADIN2} for solving~\eqref{eq::prob} to obtain $u^*$ and $\lambda^*$. 
		\item Implement $u_i^*(k)$ at subsystem~$i$, $i \in [1:\mathcal{I}]$.
		\item Reinitialize  
		\[
		u_i= (u_i^*(k+1)^\top\;\;\cdots\;\;u_i^*(k+N-1)^\top\;\;u_i^*(k+N-1)^\top)^\top
		\]
		for all $i\in[1:\mathcal{I}]$ and $\lambda = \begin{pmatrix}
		\lambda^*_2&\cdots&\lambda^*_N&\lambda^*_N
		\end{pmatrix}
		$.
		Then, set $k\leftarrow k+1$ and go to Step~1).	
		
	\end{enumerate}
	
\end{algorithm}

\section{Case Study: ALADIN vs. ADMM}\label{sec:comparison}
In this section, we compare Algorithm~\ref{alg:standard_ALADIN2} with the state-of-the-art distributed optimization algorithm ADMM. Here, we use the implementation of ADMM as proposed in~\cite{Braun2018}.

\subsection{Theoretical Comparison}
Regarding the theoretical convergence, both ALADIN and ADMM have global convergence guarantees. However, ADMM \change{only achieves} linear convergence while ALADIN converges locally with \change{a} quadratic rate. The comparison in details is listed in the Table~\ref{tab::comparison} (left). %
\begin{table}[htbp!]
	\footnotesize
	\caption{Convergence and communication comparison}
	\centering
	\renewcommand{\arraystretch}{1.7}
	\begin{tabular}{c|cc|cc}
		& \multicolumn{2}{c|}{Convergence} & \multicolumn{2}{c}{Online Communication} \\
		\cline{2-5}
		&  Local & Global & Forward & Backward \\ 
		\hline		
		ADMM~\cite{Braun2018} & Linear & Linear & $N$ & $N$ \\
		Algorithm~\ref{alg:standard_ALADIN2} &  Quadratic & $\checkmark$ & \change{$(3+n_i^{\mathrm{act}})N+2$} & $N + 1$ \\ 
	\end{tabular}
	\label{tab::comparison}
\end{table}
Here, \change{the symbol}~$\checkmark$ represents that Algorithm~\ref{alg:standard_ALADIN2} converges globally without mentioning the convergence rate. 

Table~\ref{tab::comparison} (right) compares the communication overhead per iteration using Algorithm~\ref{alg:standard_ALADIN2} and ADMM. 
In this context \emph{Forward} and \emph{Backward} are to be understood that one local agent uploads information to the CE and vice versa and~$n_i^\mathrm{act}$ is the number of active constraints at the \mbox{$i$-th} system. The table shows that the communication overhead for both Algorithm~\ref{alg:standard_ALADIN2} and ADMM is $\mathcal{O}(N)$. 
\change{In order to update $H_i$}, the CE needs to collect the information of the active Jacobian $D_i^\mathrm{act}$ for each agent $i \in [1:\mathcal{I}]$ and broadcast~$\Pi$. Thus, the forward communication overhead is $\mathcal{O}(N)$. 

\change{
Concerning the online computational efforts,
both Algorithm~\ref{alg:standard_ALADIN2} and ADMM execute two main steps per iteration: a parallelizable and a consensus step.  
The complexity of the local problems solved by the subsystems in parallel is the same for both ADMM and Algorithm~1. The main difference between these approaches is the consensus step performed by the CE. Since the update of $\Lambda^{-1}$ is optionally required in Algorithm~1, the cost of the ALADIN consensus step in worst case is $\mathcal{O}(N^2\mathcal{I})$ while ADMM only needs $\mathcal{O}(N)$. Note that the number of subsystems~$\mathcal{I}$, in practice, is much larger than the length of prediction horizon~$N$.
However, these minor differences in communication effort and computational complexity do not dominate the run time.} 
Communication between the subsystems and the CE is the most time consuming part in context of predictive control for smart grids. Speeding up the convergence progress by reducing communication rounds is the key \change{measure} to save run time in practice. 
In the following subsection Algorithm~\ref{alg:standard_ALADIN2} and ADMM are compared numerically.

\subsection{Numerical Results}
The main advantage of ALADIN compared to state-of-the-art distributed optimization algorithms such as ADMM is the local quadratic convergence\change{, cf.} Theorem~\ref{thm::LocalConvergence}. In this subsection we show the practical importance of this fact based on numerical simulations. 
\begin{table}[htbp!]
	\small
	\caption{Parameter values}
	\centering
	\renewcommand{\arraystretch}{1.5}
	\begin{tabular}{|c|c||c|c||c|c|}
		\hline
		$N$ & $24$ & $T$ & $0.5\;[\mathrm{h}]$ & \change{$\sigma_0$} & $2.4 \cdot 10^6$\\
		\hline
		\change{$\sigma_i$} & $1$& $C_i$& $2\;[\mathrm{kWh}]$& $\hat{x}_i$& $0.5\cdot C_i\;[\mathrm{kWh}]$   \\
		\hline
		$\alpha_i$ & $0.99$ & $\beta_i$ & $0.95$&$\gamma_i$ & $0.95$ \\
		\hline 
		$\ubar{u}_i$ & $-0.5\;[\mathrm{kW}]$ & $\bar{u}_i$ & $0.5\;[\mathrm{kW}]$ &  &\\
		\hline
	\end{tabular}
	\label{tab::parameters}
\end{table}

The parameter values used throughout all experiments are listed in Table~\ref{tab::parameters}. 
Since the prediction of the future net consumption is out of the scope of this paper we use data provided by an Australian grid operator~\cite{Ratnam2017}. 

\subsubsection{Open-loop comparison}

\change{In order to illustrate the numerical performance of Algorithm~\ref{alg:standard_ALADIN2}, we compare it to ADMM and an IP method for a benchmark with $\mathcal{I}=100$. In a result, each local subsystem has $48$ variables and Problem~\eqref{eq::prob} incorporates $4824$ variables in total. Here, we initialize all algorithms with zero initial guess.}

\begin{figure}[htbp!]
	\centering
	\includegraphics[width=\linewidth]{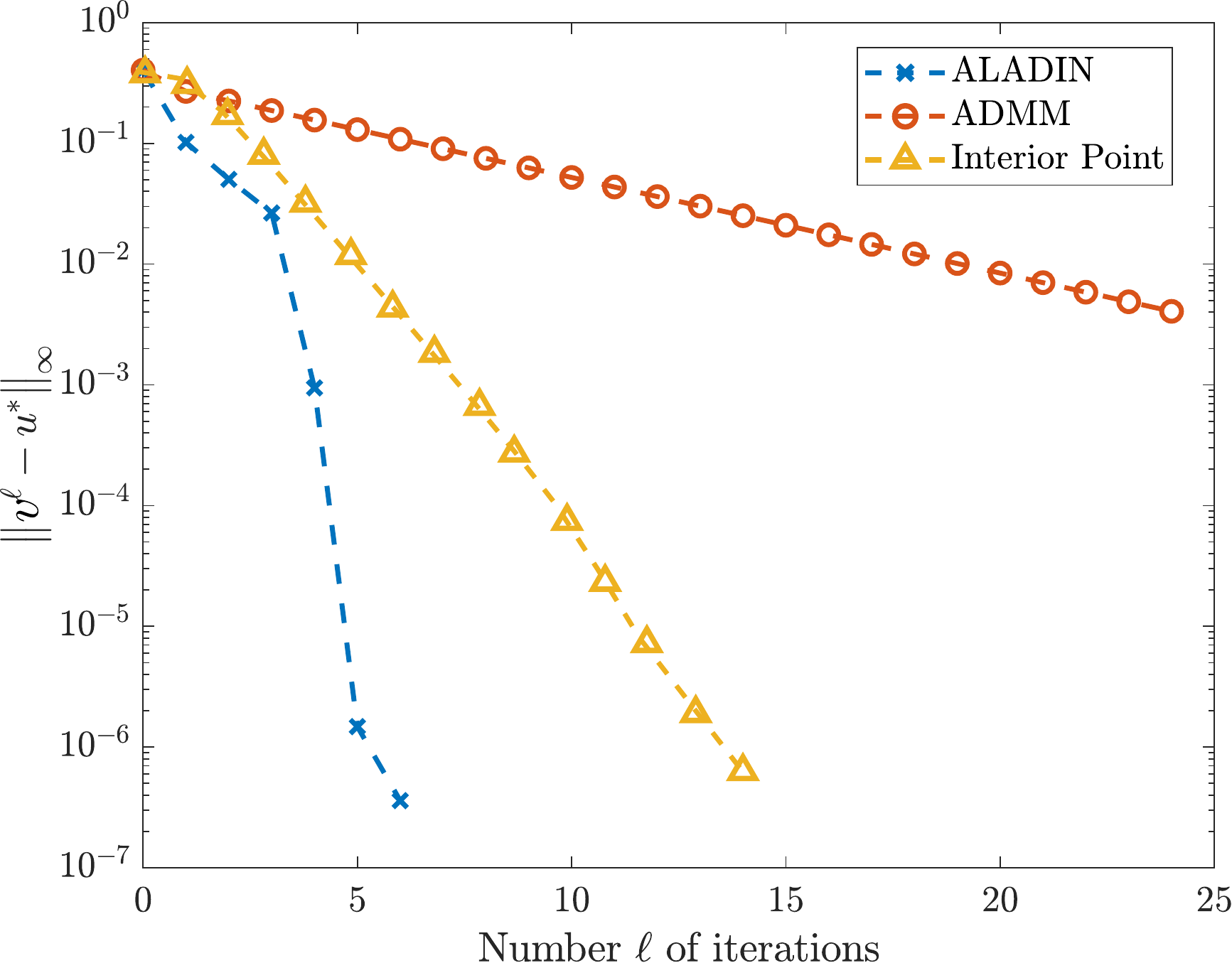}
	\caption{Convergence comparison with $\mathcal{I} = 100$ subsystems.}
	\label{fig:convergence} 
\end{figure}
\change{
The results in Figure~\ref{fig:convergence} confirm that Algorithm~\ref{alg:standard_ALADIN2} outperforms ADMM~\cite{Braun2018} and the IP method~\cite{Necoara2009}.  Here, we solved Problem~\eqref{eq::prob} for fixed time step~$k$ and initial condition and plotted the gap between the iterates~$u^\ell$ and the optimal solution~$u^*$.} 
It can be seen that ADMM needs almost $20$ iterations to achieve an accuracy of~$10^{-2}$ while ALADIN converges within six steps to an error less than~$10^{-6}$. \change{Moreover, ALADIN converges two times faster than the IP method if we choose the stop tolerance~$10^{-4}$. %
This trend can also be observed in Table~\ref{tab::cmp}, where all methods were performed based on $100$ randomly generated initial conditions. %
\begin{table}[htbp!]
	\StartChange
	\small
	\caption{Number of iterations for different methods}
	\centering
	\renewcommand{\arraystretch}{1.6}
	\begin{tabular}{cccc}
		\hline
		$\|v^\ell-u^*\|_\infty$  & Algorithm~\ref{alg:standard_ALADIN2} & ADMM & Interior Point \\ 
		\hline
		$10^{-2}$    & $4\pm2$\footnotemark[1] & $26\pm7$ & $7\pm2$\\
		$10^{-4}$    & $7\pm4$ & $64\pm15$ & $12\pm5$\\
		$10^{-6}$    & $11\pm5$ & $102\pm23$ & $18\pm6$\\ 
		\hline
	\end{tabular}
	\EndChange 
	\label{tab::cmp}
\end{table}
\footnotetext[1]{\change{This notation means that Algorithm~\ref{alg:standard_ALADIN2} takes on average $\approx4$ iterations to achieve accuracy~$10^{-2}$ while the standard deviation is $\approx2$ iterations.}}
It is worth mentioning that in every single of these randomly generated test cases Algorithm~\ref{alg:standard_ALADIN2} outperformed the other methods.
}

\change{Hence, compared to existing states-of-the-art, significantly less communication rounds between the local agents and the CE are needed in practice for Algorithm~\ref{alg:standard_ALADIN2} as discussed in Section~\ref{subsec::implementation}. Note that each iteration of the IP method requires the CE to communicate with the local agents additionally to apply a line-search routine as discussed in~\cite{Bitlisliouglu2017} even though most of its operations have been parallelized. Therefore, the IP method needs more communication rounds than ADMM. In the following, we focus on the comparison between two distributed optimization approaches, Algorithm~\ref{alg:standard_ALADIN2} and ADMM}.%

\begin{figure}[htbp!]
	\centering
	\includegraphics[width=\linewidth]{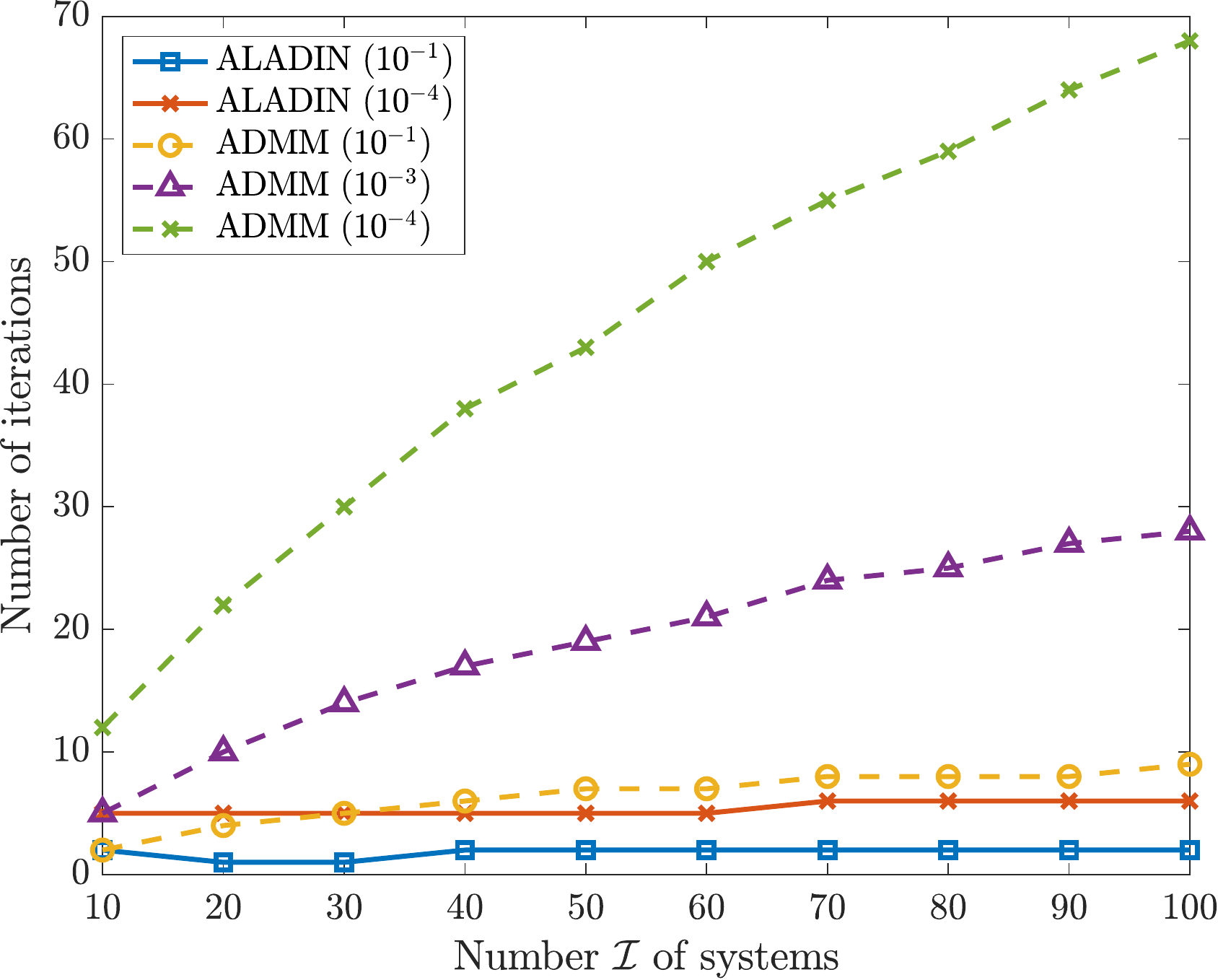}
	\caption{Impact of the number of systems on the number of iterations needed to achieve pre-defined accuracy.}
	\label{fig:iteration}
\end{figure}

In our example the complexity of the coupled problems solved within each iteration of ALADIN and ADMM, respectively, does not depend on~$\mathcal{I}$. However, the number of systems affects the number of iterations of ADMM needed to solve problem~\eqref{eq::prob}. Figure~\ref{fig:iteration} visualizes the impact of~$\mathcal{I}$ on the number of iterations of ALADIN and ADMM needed to achieve \change{different accuracies, in particular, $\{10^{-1},10^{-3},10^{-4}\}$}. It can be seen that the convergence speed of ALADIN does not depend on the number of systems while ADMM requires more iterations if the number of agents increases.

\subsubsection{Closed-loop comparison}\label{subsubsec::closed_loop}
\begin{figure}[htbp!]
	\centering
	\includegraphics[width=0.95\linewidth]{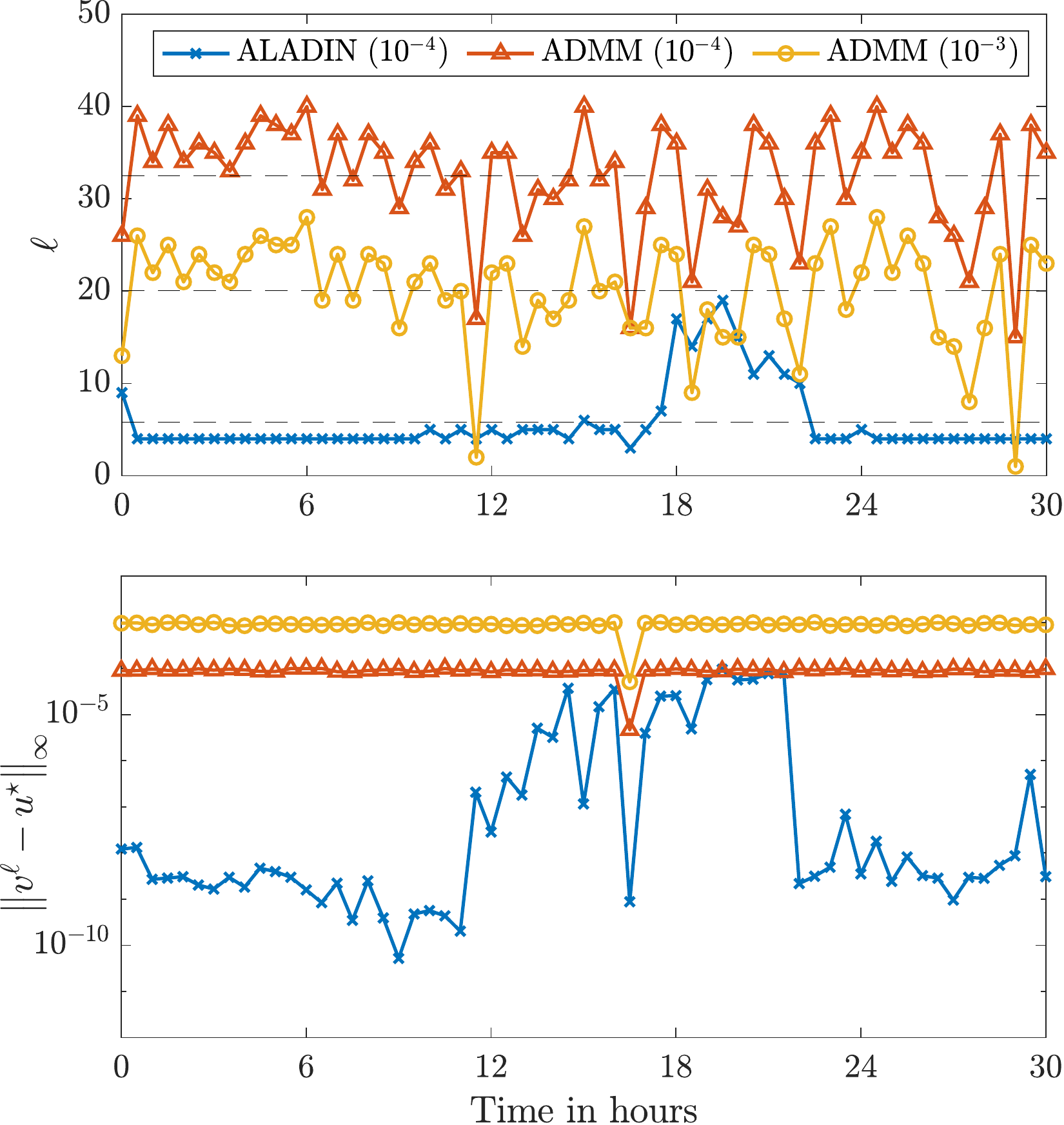}
	\caption{MPC closed loop: Number~$\ell$ of iterations needed to achieve some pre-defined accuracy (top) and actual optimization gap at iteration~$\ell$ (bottom). The dashed black lines (top) represent the respective average number of iterations.}
	\label{fig:NumIter}
\end{figure}
Figure~\ref{fig:NumIter} is dedicated to the closed-loop performance of Algorithm~\ref{alg:standard_ALADIN2} and ADMM for $30$ hours, i.e., $60$ MPC steps. More precisely, Figure~\ref{fig:NumIter} (top) compares the effort needed to achieve a pre-defined accuracy using ALADIN and ADMM by counting the executed iterations. Again the local quadratic convergence rate of ALADIN outperforms ADMM. %
\change{ALADIN needs approximately six iterations in average to achieve accuracy $\left\|v^\ell - u^*\right\|_\infty < 10^{-4}$ while ADMM needs more than~20 and 30~iterations for accuracy $10^{-3}$ and $10^{-4}$, respectively}. 
Both algorithms benefit from warm start techniques used in the implementation when $k\geq N$. Since the reference value $\change{\zeta}(k+1)$ strongly differs from $\change{\zeta}(k)$ with $k < N$, the optimal dual variable might change abruptly within consecutive MPC iterations as well. In a result, the warm start as outlined in Algorithm~\ref{alg::MPC} does not yield a reasonable performance if $k<N$. To avoid this effect, we only compare the closed-loop performance after $N$ steps. 
After that, updating $\change{\zeta}$ equals a simple shift in time enabling warm start. However, in a particular time window ($18-22$h), the warm start does not work efficiently due to the big changes of the optimal active sets. The value of $\Psi$ does not always decrease in such cases, which means that ALADIN needs more steps to detect the correct active set before achieving the local quadratic convergence progress.
Note that in some MPC iterations ADMM needs less iterations since the initial guess based on the warm start is close to the optimum. In the majority of the cases the quadratic convergence of ALADIN yields even better performance than aimed for as visualized in Figure~\ref{fig:NumIter} (bottom). 
A potential future direction to reduce the number of communication rounds further would be to prematurely stop the algorithm~\cite{Bestler2019}.

\section{Conclusions and Outlook}\label{sec:conclusion}
\change{
In this paper we proposed an optimization scheme based on ALADIN for the optimal control of locally distributed energy storages to achieve an overall goal. Thanks to its locally quadratic convergence it outperforms state-of-the-art methods such as ADMM. A numerical case study illustrates that our approach reduces the number of communication rounds required to achieve a given accuracy tremendously. 
Future research will investigate the impact of flattening the aggregated power demand on congestion in the grid.
}

\ifCLASSOPTIONcaptionsoff
  \newpage
\fi



\bibliographystyle{IEEEtran}
\bibliography{paper}
\balance

\end{document}